\documentclass[]{aa}  
\usepackage{multirow}
\usepackage{float}
\usepackage{color}
\usepackage{graphicx}
\usepackage{subcaption}
\usepackage{txfonts}

 \def\mean#1{\left< #1 \right>}
\begin{document} 
	\title{The \textit{NuSTAR} view of the Seyfert Galaxy HE 0436-4717\\}
	\author{R. Middei 
		\inst{1  \thanks{riccardo.middei@uniroma3.it}}
		\and F. Vagnetti \inst{2} \and F. Tombesi \inst{2,3,4,5}  \and S. Bianchi \inst{1} \and A. Marinucci \inst{1} \and F. Ursini \inst{6}  \and  A. Tortosa \inst{7}
	}
	\institute{Dipartimento di Matematica e Fisica, Universit\`a degli Studi Roma Tre, via della Vasca Navale 84, I-00146 Roma, Italy
\and
Dipartimento di Fisica, Universit\`a di Roma "Tor Vergata", via della Ricerca Scientifica 1, I-00133, Roma, Italy
\and
INAF Astronomical Observatory of Rome, Via Frascati 33, 00078 Monteporzio Catone, Italy
\and
Department of Astronomy, University of Maryland, College Park, MD 20742, USA
\and
NASA/Goddard Space Flight Center, Code 662, Greenbelt, MD 20771, USA
\and
 INAF-Osservatorio di astrofisica e scienza dello spazio di Bologna, Via Piero Gobetti 93/3, 40129 Bologna, Italy.
\and
INAF/Istituto di Astrofisica e Planetologie Spaziali. Via Fosso del Cavaliere - 00133 Roma - Italy
}
	
	\abstract{}
	
	\abstract
	{We present the multi epoch spectral analysis of \object{HE 0436-4717}, a bright Seyfert 1 galaxy serendipitously observed by the high energy satellite \textit{NuSTAR} four times between December 2014 and December 2015. The source flux shows modest variability within each pointing and among the four observations. Spectra are well modelled in terms of a weakly variable primary power law with constant photon index ($\Gamma$=2.01$\pm$0.08). A constant narrow \ion{Fe} K$\alpha$  emission line suggests that this feature has an origin far from the central black hole, while a broad relativistic component is not required by the data. The Compton reflection component is also constant in flux with a corresponding reflection fraction  R=0.7$^{+0.2}_{-0.3}$. The iron abundance is compatible with being Solar (A$_{\rm{Fe}}$=1.2$^{+1.4}_{-0.4}$), and a lower limit for the high energy cut-off E$_{\rm{c}}>$280 keV is obtained. Adopting a self-consistent model accounting for a primary Comptonized continuum, we obtain a lower limit for the hot corona electron temperature kT$_{\rm{e}}>$65 keV and a corresponding upper limit for the coronal optical depth of $\tau_{\rm{e}}<$1.3.
	The results of the present analysis are consistent with the locus of local Seyfert galaxies in the kT$_{\rm{e}}$-$\tau_{\rm{e}}$ and temperature-compactness diagrams.
	}
	
	\keywords{galaxies:active – galaxies:Seyfert – quasars:general – X-rays:galaxies
	}

	
	\maketitle
	\section{Introduction}
	
	Active galactic nuclei (AGN) emit in all the electro-magnetic bands, from the radio to the gamma ray domain, and this broadband phenomenology is a clue of their complex structure. According to the standard paradigm, a supermassive black hole (SMBH) lying in the innermost region of the host galaxy is sourrounded by a disc shaped inflow of matter. 
	The gravitational energy is converted into radiation and this process is responsible for the optical-UV emission of active galaxies. 
	On the other hand, X-rays are thought to be the result of an Inverse-Compton process bewteen optical-UV photons and a distribution of hot thermal electrons \citep{haar91,haar93,haar94}. This process takes place in a compact region close to the BH, the so-called hot corona. 
	While the interplay of the coronal temperature (kT$_{\rm{e}}$) and its optical depth ($\tau_{\rm{e}}$) is found to drive the AGN power law like spectral shape, the high-energy turnover is mainly a function of the coronal temperature \citep{Rybi79}. 
    Various high energy cut-offs were measured in the past \citep[e.g][]{Pero00,Nica00,DeRo02}, and since the launch of \textit{NuSTAR} an increasing number of measurements have been obtained \citep[e.g.][]{Fabi15,Fabi17,Tort18a}. 
	The radiation rising from the hot corona can be also reprocessed by the circumnuclear environment, thus additional spectral complexities such as the \ion{Fe} K$\alpha$ emission line and a Compton hump peaking at about $\sim$30 keV \citep[][]{Matt91,Geor91} is observed.
    The \ion{Fe} K$\alpha$ line is found to be narrow and likely produced by distant material \cite[e.g.][]{Cappi06,Bian07,Bian09} or to be broad. When observed, the broad component may arise from neutral and ionized iron and it can be interpreted the reflection of hard X-rays off the inner edge of an accretion disc \citep[e.g.][]{Fabi89,Reyn13}. The broadening of the lines therefore may be the result of the relativistic effects occurring close to the BH. The \ion{Fe} K$\alpha$ is often observed to be a superposition of the narrow and a broad component \citep[e.g.][]{Nand07}.\\
	\indent
	In this work we discuss the spectral analysis on four serendipitous \textit{NuSTAR} observations of HE 0436-4717 extracted from the \textit{NuSTAR} Serendipitous Survey \citep{Lans17}.
	The AGN HE 0436-4717 lies in the field of view of the pulsar PSR J0437-4715 which was the target of the observations \citep[][]{Guillot16}, and it is 4 arcmin apart. HE 0436-4717 is one of the few AGN that have been pointed by \textit{NuSTAR} in multiple epochs, and is the brightest among those serendipitously observed.
	This source is a type 1 Seyfert galaxy \citep{Vero06} lying at redshift $z$=0.053 \citep{Wiso00}, and hosting a supermassive black hole with mass $M_{\rm{BH}}=5.9\times10^{7}$ $M_\odot$, \citep{Grup10}.
	The spectral coverage of this active galactic nucleus is very peculiar since it is one of the eight AGN that have been detected in the extreme ultraviolet (EUV) band \citep{Barstow03}. Moreover, a long monitoring of $\sim$ 20 days by the \textit{Extreme UltraViolet Explorer} (EUVE) allowed to find a possible periodic variability (P=0.9 days) in the EUV \citep{Halp96,Helpern03,Leighly05}. In the X-rays, based on \textit{ASCA} and \textit{ROSAT} observations, \citet{Wang98} showed that it was possible to reproduce the HE 0436-4717 spectrum with a power law with $\Gamma\sim$2.15 and a black body with temperature 29$\pm$2 eV accounting for the soft X-ray emission. A narrow \ion {Fe} K$\alpha$ emission line was also detected in the two \textit{ASCA} observations with equivalent widths of 430$\pm$220 and 210$\pm$110 eV respectively. 
	Moreover, the authors found no spectral variability, while the source continuum increased in flux remarkably ($\sim$50\%) among the pointings (4 months apart) the 2-10 keV flux being in the range 2.9-4.4 $\times10^{-12}$ erg/cm$^{2}$/s.
	\cite{Bons15} analysed more recent \textit{XMM-Newton} and \textit{Swift} data, testing three models: partial covering absorption, blurred reflection, and
	soft Comptonization. All these scenarios were consistent with the data on a purely statistical basis.
	On the other hand, the authors argued that if the source variability and the UV emission are taken into account, the blurred reflection model provides the best self-consistent view of the data. According to this model, \citet{Bons15} found that the emission of HE 0436-4717 is due to a primary continuum that dominates over the emission from a distant neutral reflector and a blurred ionized disc reflection.	Moreover, the authors modelled the \ion{Fe} K$\alpha$ using the sum of a very broad emission line ($\sigma$=3.6$^{+3.9}_{-1.1}$ keV, EW=2.5 keV) occurring at R$_{\rm{in}}$<1.8 r$_{\rm{g}}$ and a narrow component ($\sigma$=1 eV, EW=46 eV) arising from distant and neutral material. 
	This paper is organised as follows: Sect. 2 contains the data reduction and the timing properties are discussed in Sec. 3. In Sect. 4 we focus on the spectral analysis, while in Sect. 5 the results are discussed and a summary is given.
	Furthermore, the standard cosmology \textit{$\Lambda$CDM} with H$_0$=70 km/s/Mpc, $\Omega_m$=0.27, $\Omega_\lambda$=0.73, is adopted.
	
	\begin{figure*}
		\centering
		\includegraphics[width=\linewidth]{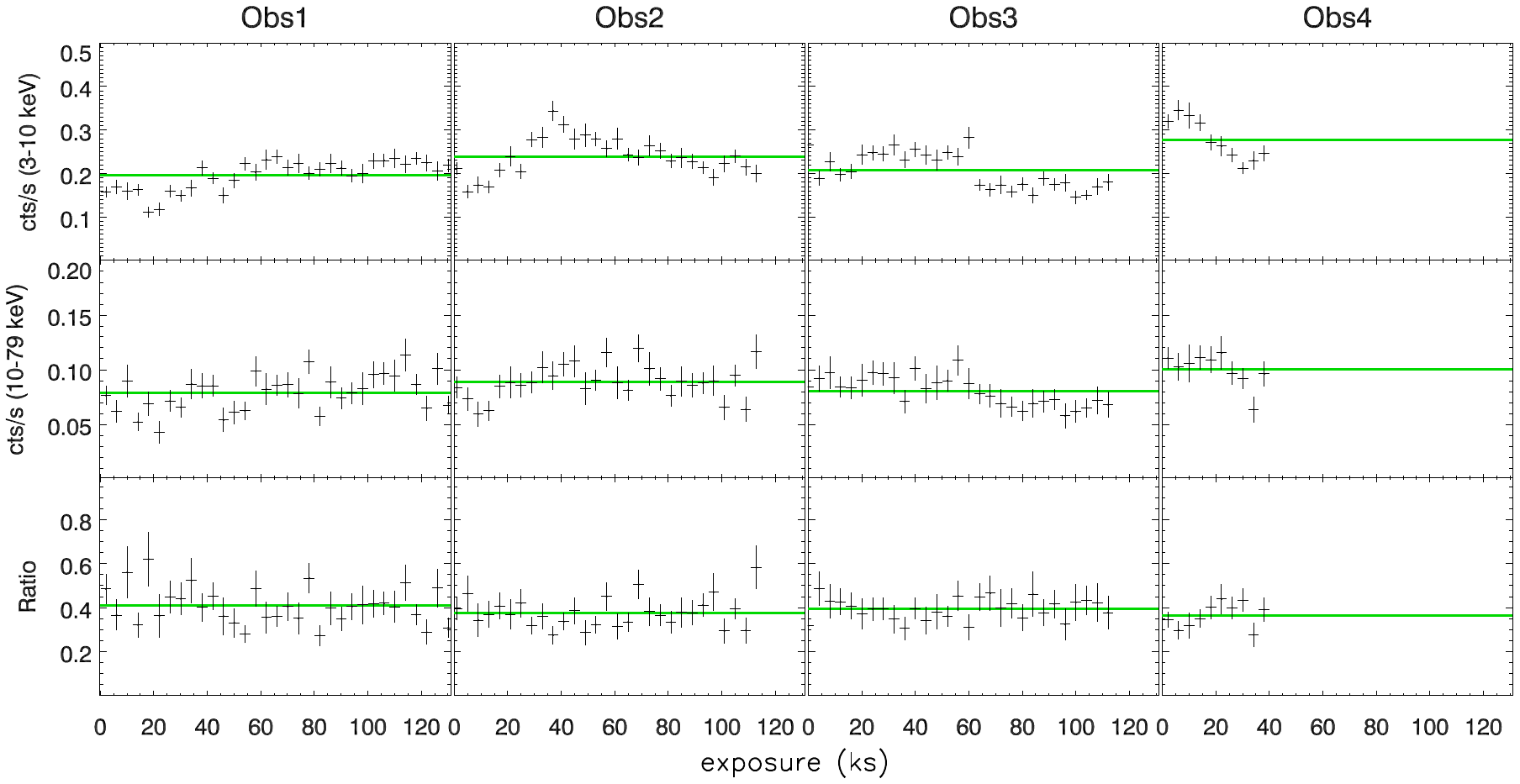}
		\caption{\small{The coadded \textit{FPMA} and \textit{FPMB} light curves are shown in the top and middle panels, in the 3-10 keV and 10-79 keV energy bands, respectively. For the various pointings, the ratios between the 10-79 keV light curves and those computed in the 3-10 keV band are shown. The adopted time binning is 4 ks for all the observations. Solid green lines account for the average count rates within each pointing. The exposures in the graph are twice as long as those reported in Tab. 1 because half of the \textit{NuSTAR} time is cut due to Earth occultations.}
		}\label{lc}
	\end{figure*}
	\section{Observations and data reduction}
	
	This analysis is based on \textit{NuSTAR} \citep{Harr13} data, and in particular, on four serendipitous observations of HE 0436-4717, reported in the \textit{NuSTAR} Serendipitous Survey \citep{Lans17}. The first three observations are separated by $\sim$1 day, while the time elapsing among the third and fourth observations is about one year, see Tab. 1.
	Therefore, long and short term flux variability and/or variation in the physical and phenomenological parameters of HE 0436-4717 can be investigated.

	\textit{NuSTAR} data were reduced using the pipeline ($nupipeline$) in the \textit{NuSTAR} Data Analysis Software (nustardas release: nustardas\_14Apr16\_v1.6.0, part of the \textit{heasoft} distribution\footnote{NuSTARDAS software guide, Perri et al. (2013), https://heasarc.gsfc.nasa.gov/docs/nustar/analysis/nustar\_swguide.pdf}), adopting the calibration database (20171204). 
		\begin{table}
		\centering
		\setlength{\tabcolsep}{2.pt}
		\caption{\small{The observation ID, the start date and the net exposure time in (ks) are reported for the \textit{NuSTAR} serendipitous observations analysed in this paper. The rates account for the average of modules \textit{FPMA} and \textit{FPMB}.}}
		\begin{tabular}{c c c c c}
			\hline
			\\
			Obs.&Obs. ID&Net rates & Net exp.&Start-date \\
			
			&&(10$^{-2}$) cts/s &ks& \\
			\hline
			\\
			1 & 30001061002 & 4.0 & 74 & 2014-12-29 \\
			\hline
			2& 30001061004 & 4.9 & 64&2014-12-31\\
			\hline
			3& 30001061006 & 4.2 &63 &2015-01-02\\
			\hline
			4& 60160197002 & 6.9 & 20 &2015-12-09\\
			\hline
		\end{tabular}
	\end{table}
	Both Focal Plane Modules A and B (\textit{FPMA/B}) on board on \textit{NuSTAR} were analysed.	We obtained the light curves and the spectra for both modules using the standard tool \textit{nuproducts}.
	To extract the source counts we used a circular region with radius of 30 arcseconds, while using a circle of the same radius, we extracted the background from a blank area close to the source.\\
	We have binned the \textit{NuSTAR} spectra in order to have a signal-to-noise (S/N) ratio greater than 3 in each spectral channel and not to over-sample the instrumental resolution by a factor larger than 2.5~.
	The obtained spectra of module A and B are in agreement with each other, their cross-normalization being within $\leq$3 per cent in all the performed fits.
	The spectra were analysed taking advantage of the standard software \textit{Xspec} 12.9.1p \citep{Arna96}.\\
	\indent In this paper, all errors in text and tables are quoted at 90\% confidence level, unless otherwise stated, and plots are in the source reference frame.
	
	\section{Temporal analysis:}
	X-ray flux variations are a hallmark of the AGN activity and they are commonly observed from years and decades \citep[e.g.][]{Vagn11,Vagn16,Midd17,Zhen17} down to hours timescales, \citep[e.g.][]{Pont12}.	Adopting the \textit{nuproducts} standard routine, we computed light curves in the 3-10 and 10-79 keV bands for HE 0436-4717, see Fig.~\ref{lc}.
	Intra-observation variability is found in the 3-10 keV light curves already at kiloseconds timescales (up to a factor $\sim$2 in observation 2), while, smaller flux variations appear in the 10-79 keV band (consistent with variability found extracting light curves in the 10-24 keV band).
	The ratios of the light curves in the 10-79 keV band and those in the 3-10 kev band are found to be compatible with being constant.	Between the different pointings, the mean counts for each of the four observations (solid line in Fig.~\ref{lc}) is found to be modestly variable. The most relevant increase of the counts ($\sim$40\%) is observed in observation four, while in the first three observations the mean of the counts has a variation of the order of 15\%. Therefore, since no strong spectral variability is found even where modest flux variations are observed, we use the average spectra of each observation to improve the spectral fitting statistics.\\
	\indent The normalized excess variance $\sigma^2_{nxs}$ \citep[e.g.][]{Nand97a,Turn99a,Vaug03,Pont12} provides a quantitative estimate of the AGN X-ray variability. This estimator can be defined as follows: $\sigma^2_{nxs}=(S^2-\sigma_{noise}^2)/\mean{f}^2~$,	
    where $f$ is the unweighted arithmetic mean flux for all the N observations, $S$ represents the variance of the flux as observed, while the mean square uncertainties of the fluxes is accounted for by $\sigma_{noise}^2$. \\
   Following this formula and computing the associated error to $\sigma^2_{nxs}$ using equation in A.1 by \cite{Pont12},
   we computed the $\sigma^2_{nxs}$ in the 3-10 keV for all the observations in 10 ks time bins, obtaining an upper limit $\sigma^2_{nxs}$<0.05. 
   Short-term
   variability has been found to be tightly correlated with the BH mass by many authors \citep[e.g.][]{Nand97,Vaug03,McHa06,Pont12}, thus adopting the relation in \cite{Pont12} for the $\sigma^2_{nxs}$ and $M_{BH}$ (see Tab. 3), we estimated a lower limit for the BH mass $M_{BH}$ > 3$\times$10$^{6}$ $M_{\odot}$, in agreement with the single-epoch measurement by \cite{Grup10}. 
\section{Spectral Analysis:}
\subsection{Phenomenological modelling}
   \begin{figure}[H]
	\centering
	\includegraphics[width=\columnwidth]{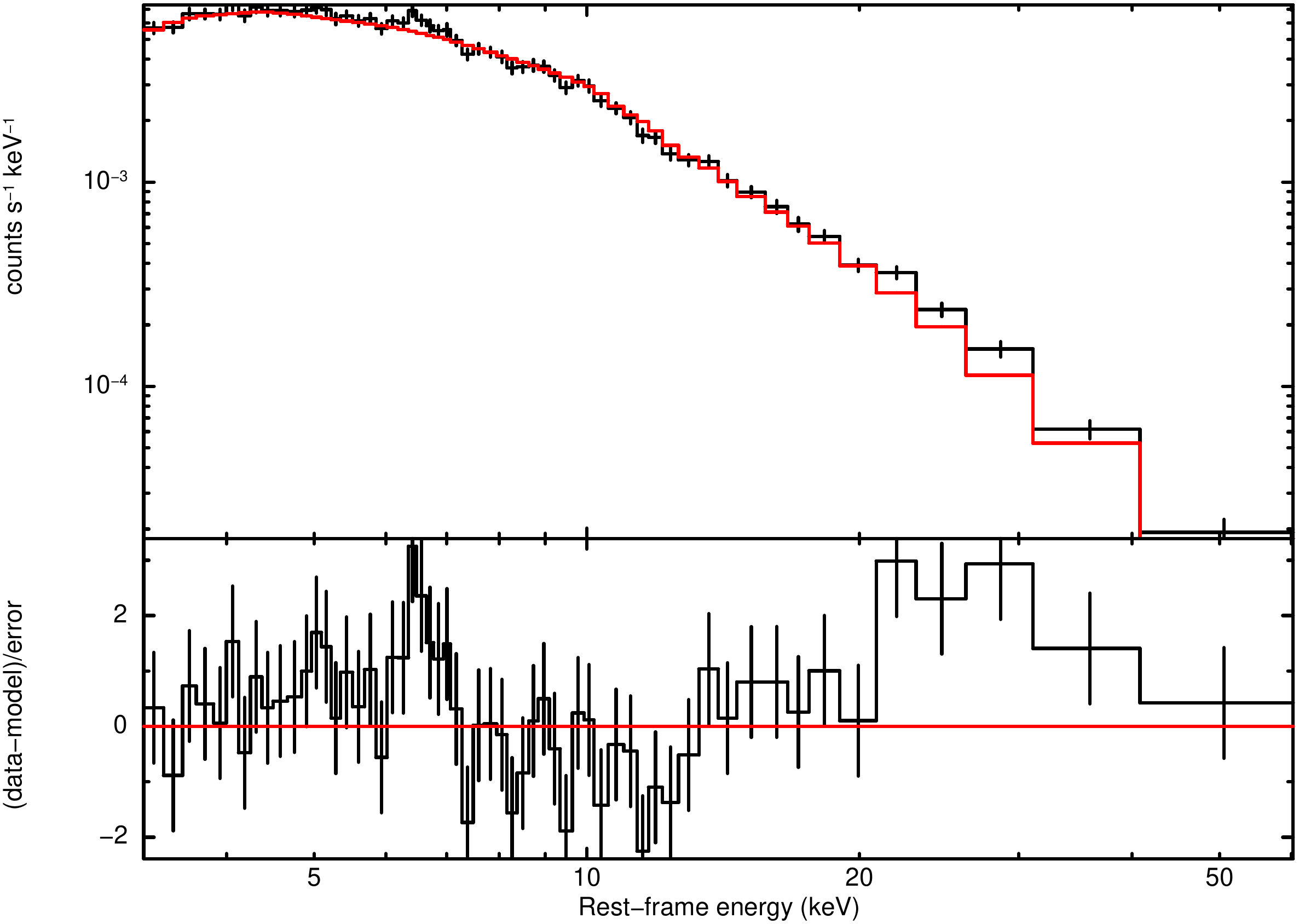}
	\caption{\small{The comparison of the data (black)
				and power law  model (red) is reported. An excess of photons is observed between 6.4 and 7 keV suggesting the presence of emission features ascribable to neutral and/or possibly ionized iron. Also a bump of unmodelled photons is found above $\sim$ 10 keV. For plotting purposes the \textit{FPMA} and \textit{FPMB} spectra of all the epochs and their residuals in terms of errors with respect to the model are displayed grouped (\textit{setplot group} in \textit{Xspec}).}}\label{res1}
\end{figure}
	As a first step, we try to reproduce the continuum emission of HE 0436-4717 with a power law absorbed by the Galactic hydrogen column \citep[N$_{\rm{H}}$=1$\times$10$^{20}$ cm$^{-2}$,][]{Kalb05}.
	In the fit, the photon index and the normalization are free to vary between the various pointings. 
	To account for the modules A and B intercalibration we use a constant set to unity for \textit{FPMA} and free to vary for \textit{FPMB}. The two modules are found in good agreement ($\leq$3\%).
	This simple model leads to a good fit ($\chi^2=460$ for 452 d.o.f) but some residuals around 6.4 keV and at energies greater than $\sim$30 keV are still present, suggesting the existence of reprocessed components (see Fig.~\ref{res1}).
	Therefore, we added a Gaussian component to account for the residuals between 6-7 keV (see Fig~\ref{res1}), obtaining the following model: \textit{const$\times$phabs$\times$(po+zgauss)}. 
	The power law shapes the primary continuum, while the \textit{zgauss} accounts for the presence of neutral or ionized emission lines. The photon index and the normalization among the different observations are untied and free to vary. For the Gaussian component we let free to vary and untied among the pointings its energy, intrinsic width ($\sigma$) and normalization. The energy of the Gaussian component is not well constrained since it is 6.50$\pm$0.15 keV. Therefore in the subsequent modelling we fix it at 6.4 keV, as for the neutral \ion{Fe} k$\alpha$. 
	In a similar fashion, the line width is found consistent with being zero in all the observations with a corresponding upper limit of 400 eV. Thus, in the forthcoming, we set its value to zero. 	This procedure yielded to a best-fit $\chi^2$=432 for 448 d.o.f and the corresponding best-fit values for the parameters are reported in Tab. 2. 
	When we fit all the observations together letting free to vary only the line normalization, a $\Delta \chi^2$=23 for 4 d.o.f. less is found.	The presence of this component is then supported by the \textit{F-test}\footnote{To reliably assess the \ion{Fe} K$\alpha$ significance via the \textit{F-test} we allowed its normalization to be negative and positive, as discussed by \cite{Protav02}.} according to which its significance is $>$ 99.9 per cent. 
	The emission line is formally detected in three over four observations, but its flux is consistent with being constant between all the pointings. 
	The average \ion{Fe} k$\alpha$ flux is 5.5$\pm$3.8 $\times10^{-6}$ ph cm$^{-2}$ $s^{-1}$ with a corresponding equivalent width of 100$\pm$10 eV.
	We also tested for the presence of a broad component of the line in our spectra.
	However, this additional broad feature is not required in terms of statistics, with a negligible $\Delta$$\chi^2$ improvement, and its normalization consistent with being zero in all the observations. 
\begin{table*}
        \setlength{\tabcolsep}{3pt}
		\centering 
		\caption{\small{The best-fit values of the parameters for all the models tested in this 3-79 keV band analysis. For each model we have accounted for the \textit{FPMA/B} intercalibration constant and the Galactic hydrogen column.
		}}
		\begin{tabular}{c c c c c c c c}
		\hline
\multicolumn{1}{c}{\multirow{5}{4cm}{Model: \textit{POWER LAW}\\$\chi^2$= 432 for 448 d.o.f.}}&&&&&\\
&Obs &  $\Gamma$ & N$_{\rm{po}}$~(10$^{-3}$)&  ~N$_{\rm{\ion Fe~K\alpha}}$~(10$^{-6}$)& EW &Flux$_{\rm{3-10}}$ (10$^{-12}$)&\\		
&& &ph/keV/cm$^2$/s& ph/cm$^{2}$/s &eV&erg/cm$^2$/s&\\	
			&1  & $1.84\pm0.05$ & $1.1\pm0.1$ & $3.1\pm2.5$ &$70^{+80}_{-60}$&2.9$^{+1.1}_{-1.0}$&\\
			\\
			&2  & $1.84\pm0.04$ & $1.3\pm0.1$ & $6.2\pm3.0$ &$130^{+70}_{-90}$&3.0$^{+0.9}_{-1.4}$&\\
			\\
			&3  & $1.84\pm0.05$ & $1.1\pm0.1$ & <4.1 &<130&2.5$^{+1.0}_{-1.5}$&\\
			\\
			&4  & $1.87\pm0.07$ & $1.6\pm0.2$ & $7.2\pm5.1$ &$110^{+120}_{-110}$&3.9$^{+1.0}_{-1.7}$&\\
			\hline
			\hline
\multicolumn{1}{c}{\multirow{5}{4cm}{Model: \textit{PEXRAV}\\$\chi^2$= 404 for 440 d.o.f.}}&&&&&\\
&Obs&$\Gamma$&E$_{\rm{c}}$&R$_{\rm{frac}}$& N$_{\rm{pexrav}}$~(10$^{-3})$&N$_{\rm{\ion Fe~K\alpha}}$~(10$^{-6})$&\\
&& & keV& &ph/keV/cm$^2$/s& ph/cm$^{2}$/s&\\
			\\
			&1  & $2.07^{+0.06}_{-0.03}$ & >100 & $1.3^{+0.7}_{-0.7}$ &$1.5^{+0.1}_{-0.1}$&$2.2\pm2.0$&\\
			\\
			&2  & $1.97^{+0.08}_{-0.03}$ & >140 & $0.6^{+0.5}_{-0.4}$ &$1.6^{+0.2}_{-0.1}$&$4.7\pm2.4$&\\
			\\
			&3  & $2.08^{+0.07}_{-0.03}$ & >130 & $1.5^{+0.7}_{-0.8}$ &$1.5^{+0.3}_{-0.2}$&<2.61&\\
			\\
			&4  & $2.02^{+0.08}_{-0.12}$ & >60 & <1.2 &$1.9^{+0.3}_{-0.1}$&$7.2\pm4.4$&\\
			\hline
			\hline
\multicolumn{1}{c}{\multirow{5}{4cm}{Model: \textit{PEXMON}\\$\chi^2$=407 for 444 d.o.f. \\A$_{\rm{Fe}}$=1.1$^{+1.3}_{-0.3}$}}&&&&&\\
&Obs&$\Gamma$&E$_{\rm{c}}$ &R$_{\rm{frac}}$& N$_{\rm{pexmon}}$~(10$^{-3}$)&&\\
&& & keV& &ph/keV/cm$^2$/s&&\\
&1  & $2.02^{+0.09}_{-0.09}$ & >150 & 0.7$^{+0.4}_{-0.3}$ &$1.4^{+0.2}_{-0.2}$&&\\
\\
&2  & $2.02^{+0.08}_{-0.12}$ & >105 & 0.7$^{+0.4}_{-0.3}$ &$1.7^{+0.3}_{-0.2}$&&\\
\\
&3  & $1.97^{+0.08}_{-0.10}$ & >130 & 0.4$^{+0.4}_{-0.3}$ &$1.4^{+0.2}_{-0.2}$&&\\
\\
&4  & $2.08^{+0.13}_{-0.12}$ & >95 & 0.8$^{+0.7}_{-0.5}$ &$2.1^{+0.4}_{-0.4}$&&\\
\hline
\hline
\multicolumn{1}{c}{\multirow{5}{4cm}{Model: \textit{MYTORUS}\\$\chi^2$= 404 for 444 d.o.f.}}&&&&&\\
&Obs. & $\Gamma$ &  N$_{\rm{po}}$ ~(10$^{-3}$)&N$_{\rm{myTorus}}$/N$_{\rm{po}}$& N$_{\rm{H}}$~(10$^{24}$)&&\\
&& &ph/keV/cm$^{2}$/s&& cm$^{-2}$&&\\
&1 &2.05$^{+0.10}_{-0.09}$&1.5$^{+0.2}_{-0.2}$&1.8$^{+0.7}_{-0.6}$&>1.6&&\\
\\
&2 &1.96$^{+0.06}_{-0.07}$&1.6$^{+0.2}_{-0.1}$&1.6$^{+0.6}_{-0.5}$&>2.4&&\\
\\
&3 &1.99$^{+0.9}_{-0.9}$&1.4$^{+0.6}_{-0.4}$&1.2$^{+0.6}_{-0.5}$&>1.8&&\\
\\
&4&2.08$^{+0.12}_{-0.12}$&2.1$^{+0.6}_{-0.4}$&2.3$^{+1.1}_{-0.9}$&>2.6&&\\
\hline
\hline
\multicolumn{1}{c}{\multirow{5}{4cm}{Model:\textit{XILLVER}\\$\chi^2$= 404 for 443 d.o.f.\\A$_{\rm{Fe}}$=1.2$^{+1.4}_{-0.4}$}}&&&&&\\
&Obs. & $\Gamma$ &  E$_{\rm{c}}$&R$_{\rm{frac}}$ & N$_{\rm{xi}}$~(10$^{-5}$)&kT$_{\rm{e}}$& $\tau_{\rm{e}}$\\
&&&keV&&ph/keV/cm$^{2}$/s&keV&\\
&1 &1.98$^{+0.13}_{-0.09}$&>140&0.6$^{+0.5}_{-0.3}$&2.3$^{+0.2}_{-0.4}$&>30&<2.4\\
\\
&2 &2.00$^{+0.12}_{-0.09}$&>115&0.7$^{+0.4}_{-0.3}$&2.9$^{+0.2}_{-0.7}$&>30&<2.4\\
\\
&3 &1.94$^{+0.16}_{-0.14}$&>110&0.4$^{+0.4}_{-0.2}$&2.5$^{+0.2}_{-0.4}$&>20&<3.3\\
\\
&4 &2.07$^{+0.16}_{-0.16}$&>120&0.8$^{+0.8}_{-0.5}$&3.0$^{+0.4}_{-0.7}$&>30&<2.2\\
\hline
\hline
		\end{tabular}
		\end{table*}
	\\
\indent The primary photon index is found to be constant among the different pointings, while weak variability is observed in the primary continuum normalization.
The unmodelled photons above 10 keV in Fig.~\ref{res1} indicate that part of the emission of \object{HE 0436-4717} is due to reflection of the primary continuum, thus we replace in our best-fit model the power law with \textit{pexrav} \citep{Magd95}.	
In this new model (\textit{const$\times$phabs$\times$(pexrav+zgauss)}),  \textit{pexrav} reproduces the power law like primary continuum with its associated reflected component, while the Gaussian line accounts for the \ion{Fe} K$\alpha$.
In the fitting procedure the iron abundance is frozen to the Solar value for all the observations, while the photon index, the normalization and the reflection fraction are free to vary between the pointings. We also let the high energy cut-off free to vary and untied among the observations. 
The adoption of this model yields to a best-fit of $\chi^2$=404 for 440 d.o.f. for which we report the best-fit values in Tab. 2, second panel.

Allowing for a reflection hump, we find that the photon index is compatible with being constant, and its best-fit values appear steeper than those previously obtained using a simple power law.
Within the errors, the reflection fraction is constant between the pointings, and the \textit{pexrav} normalization exhibits modest variations.
For the high energy cut-off only lower limits are found, see second panel in Tab. 2.
We further test the reflected Compton component using the following model: \textit{const$\times$phabs$\times$(cutoffpl+pexrav)}.
The cut-off power law  (\textit{cutoffpl}) models the primary continuum, while \textit{pexrav} shapes the reflected component only. The photon index and high energy cut-off are tied between the components and free to vary. Both the primary and reflected component normalizations are free to vary and untied. 
This model yields a best-fit of $\chi^2$=406 for 447 d.o.f, and the best-fit parameters are equivalent within the errors with those in the  second panel of Tab. 2.
We therefore fit the \textit{NuSTAR} data tying the normalization of the reflected component between the observations. The obtained fit ($\chi^2$=410 for 450 d.o.f.) is statistically equivalent to the previous one, thus a constant normalization of the reflected component is found N$_{\rm{refl}}$=1.5$\pm$0.5$\times$10$^{-3}$ ph/keV/cm$^{2}$/s with a corresponding constant flux F$_{\rm{20-40 keV}}$=2.6$\times$10$^{-12}$ erg/cm$^2$/s.
\subsection{Physical modelling}
\begin{figure*}
		\includegraphics[]{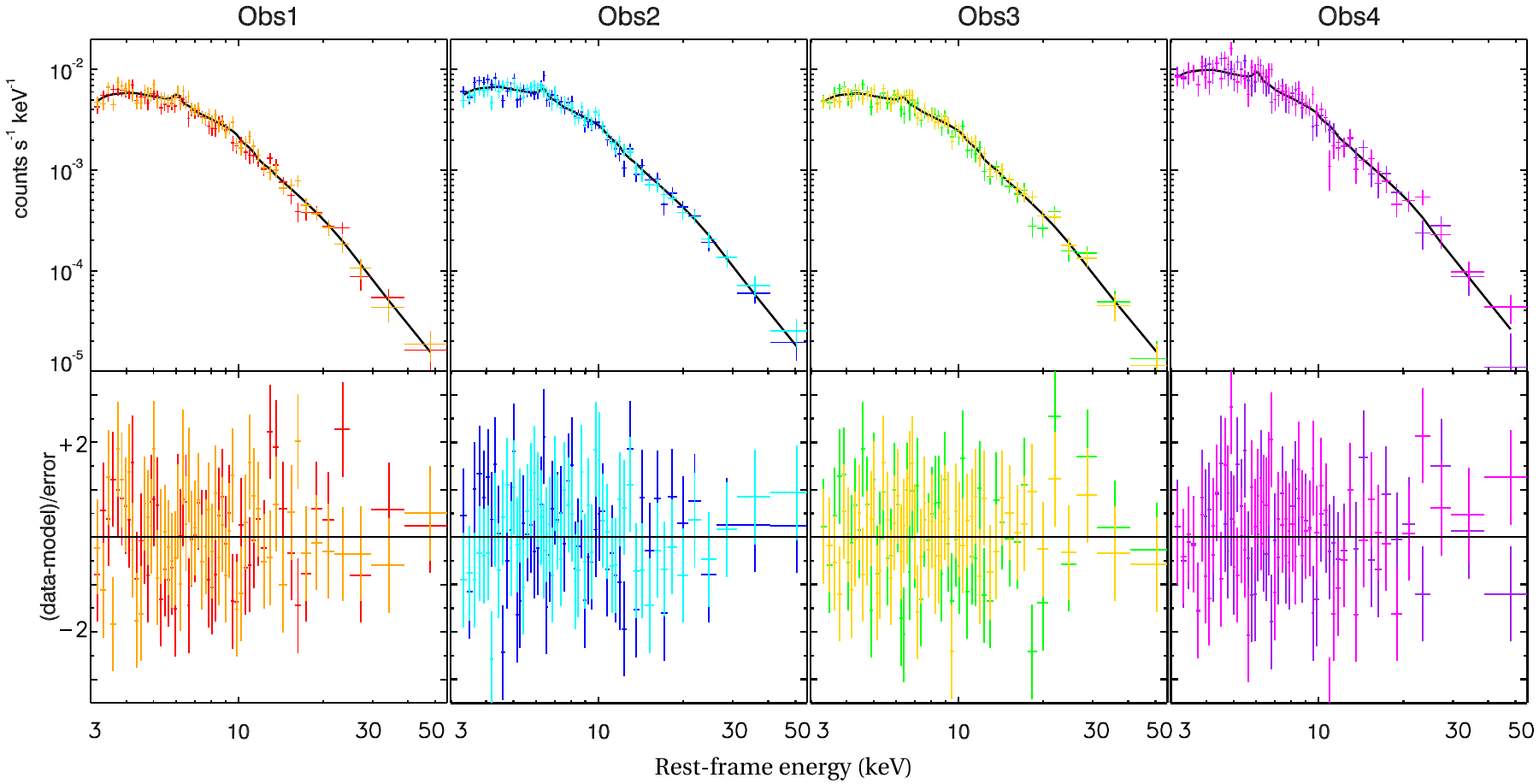}
		\caption{The best-fit model \textit{const$\times$phabs$\times$xillver} to the \textit{NuSTAR} data is displayed for each observation. In the x-axis the energy is reported in keV and different colors are used to represent the \textit{NuSTAR} module A and B spectra.}\label{bestxi}
\end{figure*}

The narrow and constant \ion{Fe} K$\alpha$ suggests that the origin of the reprocessed emission of HE 0436-4717 is far from the central BH.
However, the geometrical configuration of the reflecting material is unknown, then we tested few models 
to account for different geometries.
At first we have tried to reproduce the data set adopting \textit{pexmon} \citep[][]{Nand07}.
\indent \textit{Pexmon} combines \textit{pexrav} with self-consistently generated Fe and Ni emission lines. To fit the data with \textit{pexmon} we adopt the same procedure used for testing \textit{pexrav}, thus we let free to vary and untied between the observations the photon index, the high energy cut-off, the reflection fraction and the normalization.
Moreover, we fit the iron abundance A$_{\rm{Fe}}$ tying it among the pointings. The obtained best-fit ($\chi^2$= 407 for 444 d.o.f.) is statistically equivalent to the one in which \textit{pexrav} was adopted, and the best-fit values of the parameters are compatible each other, see Tab. 2, third panel.
Adopting this model that self-consistently accounts for the fluorescence emission lines, we estimate the iron abundance to be A$_{\rm{Fe}}$=1.1$^{+1.3}_{-0.3}$.\\
\indent As an additional test, we have fitted our data set adopting \textit{MYTORUS} \citep{Yaqo12}. Through this model we further test the origin of the HE 0436-4717 reprocessed emission. In fact, \textit{MYTORUS} accounts for a narrow \ion{Fe} K$\alpha$ and its accompanying reflection component by a Compton-thick toroidal material.
We assumed a power-law like illuminating continuum. Both the $\Gamma$ and normalization of the primary emission are free to vary and untied between the pointings. At first, we have used the \textit{MYTORUS} tables accounting for the emission lines and scattering untied and free to vary. However the two table normalizations were consistent with each other. Thus we kept them tied together accordingly with the coupled reprocessor solution \citep{Yaqo12}. Then, for each observation, we tied the underlying continuum $\Gamma$ with the one characterizing the reflected emission. Finally, we performed the fit tying the normalizations of the primary continuum and reprocessed component and adding a further constant to account for their mutual weights. This procedure leads to a best-fit characterized by $\chi^2$=404 for 444 d.o.f, and in the fourth panel of Tab. 2 we report the corresponding best-fit values. We notice that the constant accounting for the relative normalizations
of the primary emission and reflected one (N$_{\rm{myTorus}}$/N$_{\rm{po}}$) has fairly high values. These may suggest a larger covering factor with respect to the default one in \textit{MYTORUS}. However, the interpretation of this constant is not trivial since it embodies different degenerate information about the chemical abundances and the covering factor itself, \citep[see][]{Yaqo12}.
Moreover, from the fit, only lower limits are obtained for the column density of the reflectors (1.6-2.4 $\times$10$^{24}$ cm$^{-2}$).\\
\indent Finally we tested \textit{Xillver} \citep[][]{Garc10,Garc13}, a model that reproduces the primary continuum and  reflection off an accretion disc.
\textit{Xillver} assumes a cut-off power law for the primary emission, and it is commonly used to model reflection from distant material \citep[e.g.][]{Park16}. We performed the fit letting free to vary the photon index, the reflection fraction, the normalization and the high energy cut-off for each observation. The iron abundance was free to vary but tied among the observations. At first we also fitted the ionization parameter $\xi$, but, since no improvements were found during the fitting procedure, we fixed its value to the lowest allowed by the model (log$\xi=$0), close to neutral matter.
The obtained best-fit parameters can be found in the fifth panel of Tab. 2, while the best-fit model for all the observations is shown in Fig.~\ref{bestxi}.

    \begin{figure}[]
	\includegraphics[width=\columnwidth]{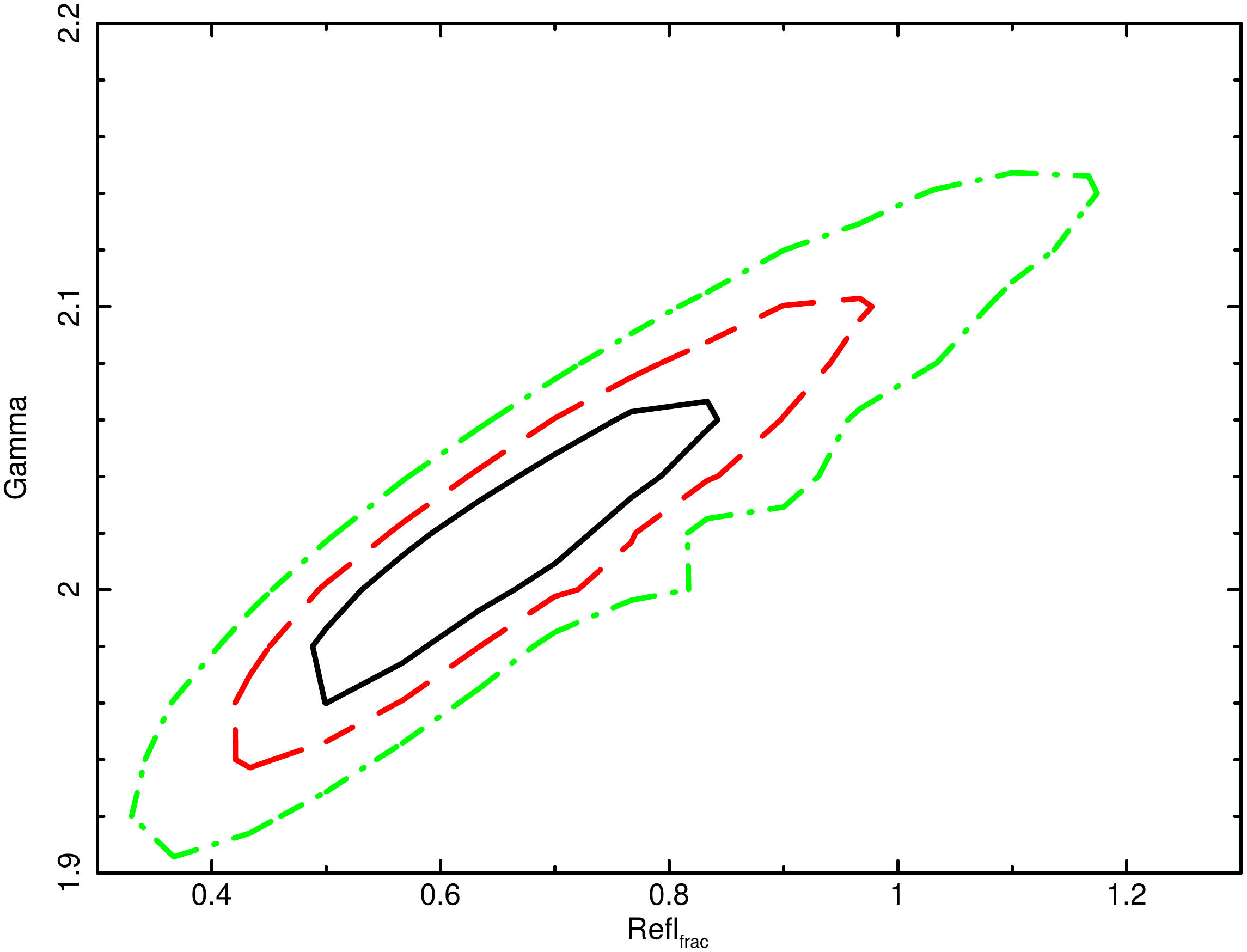}
	\includegraphics[width=\columnwidth]{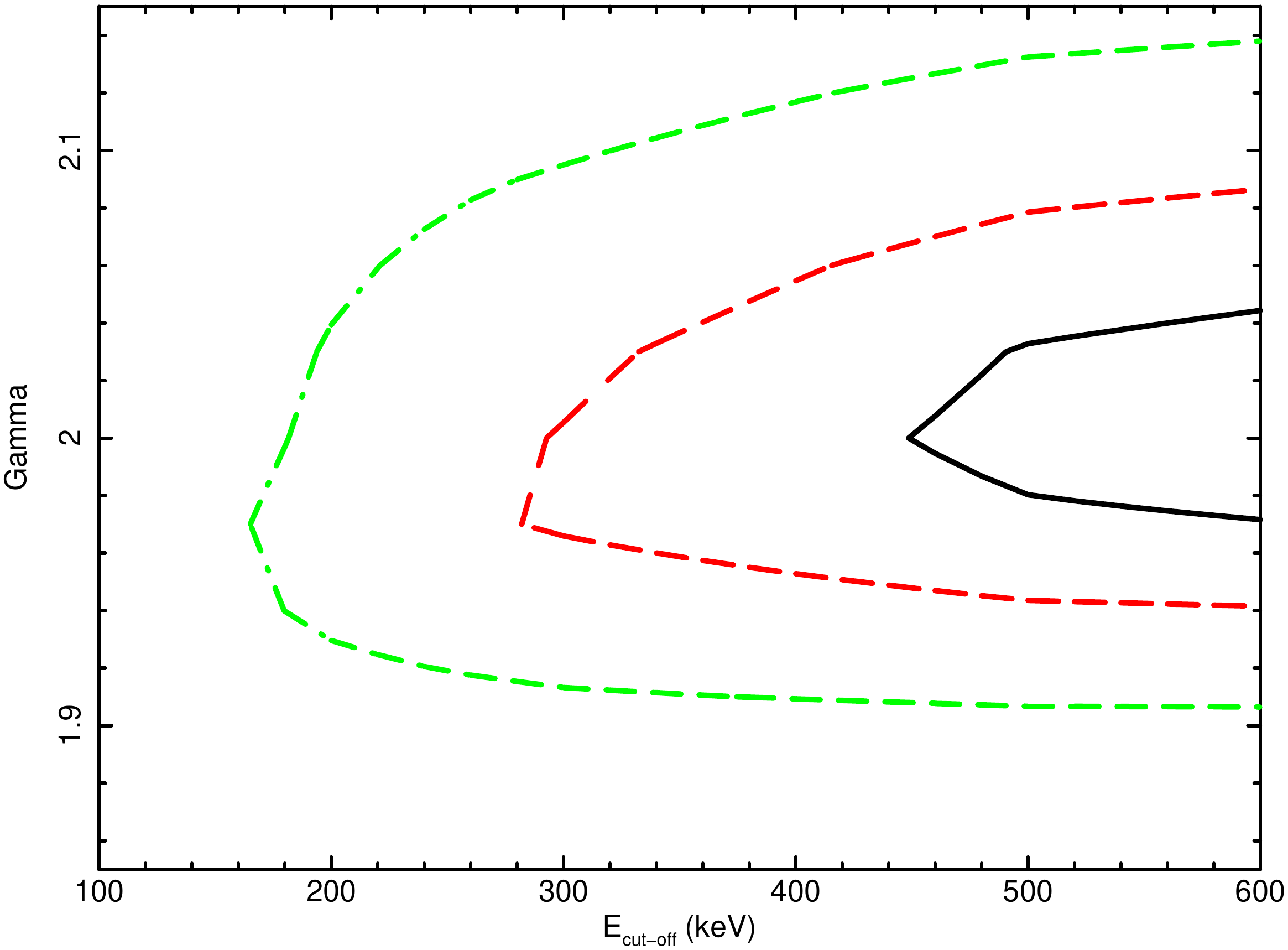}
	\caption{\small{The 68\%, 90\% and 99\% contours in black, red, green respectively computed adopting  \textit{const$\times$phabs$\times$xillver}. Top panel: The photon index and the reflection fraction are shown. Bottom panel: Contours for the photon index and the high energy cut-off.}}\label{cont}
\end{figure}
    As found adopting pexmon, the photon index is compatible with being constant, and, similarly, the iron abundance is found to be A$_{\rm{Fe}}$=1.2$^{+1.4}_{-0.4}$.
    Beside a marginally variable underlying continuum, most of the components have a constant behaviour, thus we try to fit the data tying the photon index and the high energy cut-off between the various pointings. The reflection fraction and normalization of the primary continuum are untied and free to vary among the observations.  
    Following this procedure we obtained a best-fit statistically equivalent with the previous one just discussed ($\chi^2$=406 for 449 d.o.f. \textit{vs} $\chi^2$=404 for 443 d.o.f~).
    The best-fit value for the photon index is $\Gamma$=2.01$\pm$0.08 while the obtained lower limit for the high energy cut-off is E$_{\rm{cut}}>$280 keV.
    We therefore tried to better constrain the reflection fraction R letting its value free to vary but tied among the observations. This procedure does not change the quality of the fit, but allow us to estimate the averaged value of the reflection fraction to be R=0.7$^{+0.2}_{-0.3}$.  
    We then compute the contour plots for the parameters that are shown in Fig~\ref{cont}.\\
    \indent Comptonization is widely accepted to be the origin of the X-ray emission in AGN, thus, substituting \textit{xillver} with \textit{xillvercp} \citep[][]{Garc10,Garc13}, we tried to investigate the physical properties of the HE0436-4717 hot corona. 
    This model differs from \textit{xillver} because the primary continuum is shaped by thermal Comptonization through the \textit{nthcomp} model \citep{Zdzi96}. We perform the fit tying the parameters among the observations, and we let free to vary only the hot electrons temperature. In terms of statistics this fit ($\chi^2$= for 405 d.o.f. 443) is compatible with the previous one in which \textit{xillver} was adopted. 
    For sake of simplicity, in the panel referring to \textit{xillver} in Tab. 2 we only report the obtained values for the hot electron temperature kT$_{\rm{e}}$, as the other parameters are compatible with those already obtained using \textit{xillver} within error bars.
    Furthermore, assuming a spherical geometry and using the \textit{nthcomp} internal routine for the Thomson optical depth $\tau_{\rm{e}}$, we obtained upper limits for the coronal optical depth of He 0437-4717. These upper limits are reported in the fifth panel of Tab. 2.
    Moreover, we fit again the data tying the kT$_{\rm{e}}$ among the observations ($\chi^2$= for 408 d.o.f. 452). This procedure leads to a lower limit for the corona temperature: kT$_{\rm{e}}$>65 KeV.
    Again, under the assumption of a spherical Comptonizing medium, we computed a corresponding optical depth $\tau_{\rm{e}}$< 1.3. Even though the analysed \textit{NuSTAR} data did not require any relativistic component, this feature seems to be needed by the \textit{XMM-Newton} data. 
    Thus we have tested for the presence of a broader and relativistic component adding \textit{relxill} to our best-fit model.
    Since the parameters are not well constrained, we set \textit{relxill} according to the values reported by \citet{Bons15}: $\Gamma$=2.14, R$_{\rm{in}}$=1.8 r$_{\rm{g}}$, A$_{\rm{Fe}}$=0.36, inclination $\theta$=43$^\circ$. However, adding \textit{relxill} yields a fit of $\chi^2$=406 for 445 d.o.f., thus, on statistical basis, these \textit{NuSTAR} data do not require a relativistic component.\\

\section{Discussion}
The \textit{NuSTAR} high sensitivity above 10 keV makes it suitable for investigating the physical conditions of the AGN coronae. However, the lack of sufficient statistics limits comparative studies concerning the AGN coronal region. In fact, at present, the largest sample of these sources analysed taking advantage of \textit{NuSTAR} data counts few AGN only (<20), \citep[e.g.][]{Fabi17,Tort18a}. In this framework, the analysis of these multi epoch \textit{NuSTAR} observations adds information about the coronal parameter of this particular Seyfert galaxy, enlarging at the same time the number of AGN analysed thanks to \textit{NuSTAR}.\\
\indent The 3-79 keV HE 0436-4717 \textit{NuSTAR} spectra are found to be consistent with being the superposition of two spectral components, a persistent and weakly variable primary emission and a narrow iron K$\alpha$ with its associated distant reflection continuum. The primary continuum can be described by a power law with photon index $\Gamma$=2.01$\pm$0.08 and a lower limit E$_{\rm{c}}$>280 keV for the cut-off energy. Moreover, a neutral \ion{Fe} K$\alpha$ emission line is present in three over four observations and it is found to be narrow. The reflected component is found compatible with being constant in flux, and the high energy emission of HE 0436-4717 is in agreement with a scenario in which this reprocessed emission arises from neutral material far from the central engine. Therefore, we have tested a few models that account for different geometrical scenarios. Disc reflection provides a good fit to the data, and, similarly, reprocessed emission by toroidal matter with N$_{\rm{H}}\gtrsim2\times$10$^{24}$ cm$^{-2}$ is statistically supported.\\
The \textit{NuSTAR} data analysed in this work do not require a broad line, consistent with \citet{Wang98} who found the line to be narrow and likely due to distant reflection. However, \textit{XMM-Newton} data requires a broad component \citep{Bons15}. This discrepancy can be ascribed to the poorer spectral resolution of \textit{NuSTAR} with respect to \textit{XMM-Newton}, since part of the broad component flux might be absorbed by the narrower feature. Moreover, we are further limited in testing a complete blurred reflection scenario because any soft-excess would occur outside the \textit{NuSTAR} operating band.
Adopting different models, the source iron abundance is found within the errors compatible with being Solar ($A_{\rm{Fe}}$=1.2$^{+1.4}_{-0.4}$).
 HE 0436-4717 shows modest flux variations within each observation while the average of the counts exhibits a more constant behaviour between the \textit{NuSTAR} pointings. The largest flux variations are measured on timescales shorter than a day. Past flux measures in the 2-10 keV band performed using \textit{ASCA} and \textit{XMM-Newton} data, \citet{Wang98} and \cite{Bons15} respectively, are compatible (3.3-4.6$\times$10$^{-12}$ erg/cm$^{2}$/s) with the flux observed during the \textit{NuSTAR} pointings.
 We estimated the 2-10 keV luminosity to be L$_{\rm{2-10~keV}}$=3$\pm$0.5$\times$10$^{43}$ erg/s. Adopting the proper bolometric correction from \cite{Marconi04}, we found the bolometric luminosity of HE 0436-4717 to be L$_{\rm{bol}}$=7$\times$10$^{44}$ erg/s.
 Considering a M$_{\rm{BH}}$=5.9$\times$10$^{7}$M$_\sun$ \citep{Grup10}, HE 0436-4717 has an Eddington ratio of $L/L_{\rm{Edd}}\sim$ 0.09.\\
 \indent This spectral analysis reveals that HE 0436-4717 did not experiment any spectral variation, and this is compatible with what was found in previous works. In fact, \citet{Wang98} measured a photon index $\Gamma$=2.15$\pm$0.04, while \cite{Bons15} obtained $\Gamma$=2.12$\pm$0.02. 
 Our \textit{NuSTAR} measures are compatible with previous estimates, thus no evidence for long term spectral variability is found.\\\indent     
 The adoption of realististic models including Comptonization allowed us
 to study the coronal physics of different AGN \citep[e.g.][]{Petr13,Midd18,Ursini18}, thus we included in our analysis \textit{nthcomp} to investigate the coronal properties of HE 0436-4717.
 Accounting for Comptonization and tying the electrons temperature among the various pointings, a lower limit for the coronal temperature is found: kT$_{\rm{e}}>$65 keV.
 From this value we derived an upper limit for the electron optical depth of $\tau_{\rm{e}}<$1.3~.
 Our values for the $\tau_{\rm{e}}$ and kT$_{\rm{e}}$ are in agreement with what is expected for an optically thin and hot medium responsible for the AGN hard X-ray emission. \cite{Tort18a} present results on the hot corona parameters of 19 AGN measured with NuSTAR, and, in particular, the authors discuss various relations between phenomenological parameters and physical ones. The HE 0436-4717 $\tau_{\rm{e}}$ and kT$_{\rm{e}}$ values are in perfect agreement with the strong anti-correlation they found for the optical depth and the coronal temperature of the 19 Seyfert galaxies.\\
 \indent Moreover, the estimated temperature kT$_{\rm{e}}$ can be used to investigate how HE 0436-4717 behaves on the compactness-temperature diagram \citep[][and references therein]{Fabi15,Fabi17}.
 These two parameters are defined as follows: $\Theta_{\rm{e}}$=kT$_{\rm{e}}$/m$_{\rm{e}}$c$^2$ and $l$=$\frac{L}{R}$$\times\sigma_{\rm{T}}$/m$_{\rm{e}}$c$^3$.
 The first equation accounts for the coronal electron temperature normalised by the rest-frame energy of the electrons, while the second one is used to define the dimensionless compactness parameter \citep[][]{Fabi15}. In this latter formula $L$ is the luminosity, and R is the radius of coronal (assumed to be spherical). For HE 0436-4717 $\Theta_{\rm{e}}>0.13$ is obtained. Following \citet{Fabi15}, we computed the luminosity extrapolating its value to the 0.1-200 keV band, $L$=1.4$\times$10$^{44}$. Since R is not measured, we assume a value of 10 r$_{\rm{g}}$. We then compute the compactness of the HE 0436-4717 to be $l$=230~(R$_{10}$)$^{-1}$, where R$_{10}$ is just the ratio between the radius and 10 r$_{\rm{g}}$. In the $\Theta_{\rm{e}}-l$ diagram by \citet{Fabi15}, HE 0436-4717 lies, as does the bulk of the sample analysed by the authors, below the forbidden runaway pair production line. This supports that AGN coronae are hot and radiatively compact.  
 The flux of the reflected component is found to be constant between the observations, and a corresponding averaged reflection fraction R=0.7$^{+0.2}_{-0.3}$ is obtained. This latter value is compatible with the bulk of the measurements commonly found for Seyfert galaxies \citep[e.g.][]{Pero02,Ricci17}.\\

    \section{Summary}
     This paper focuses on the spectral properties of the Seyfert galaxy HE 0436-4717, and it is based on the analysis of four serendipitous \textit{NuSTAR} observations performed from December 2014 to December 2015.
The main results of this analysis are:
\begin{itemize}
	\item Modest flux variability is observed within the various \textit{NuSTAR} observations at few kilo seconds timescales.  The average count rate of each epoch in the 3-10 and 10-79 keV bands is only weakly variable.
	Moreover, we have quantified the source variability computing the normalised excess variance for this source. We obtained an upper limit for this estimator, $\sigma_{\rm{NXS}}<$0.05~. We then converted this value into a lower limit for the BH mass obtaining M$_{\rm{BH}}>$3$\times$10$^{6}$M$_{\odot}$, in agreement with the single epoch measure by \cite{Grup10}.
	\item A power law like spectrum with a corresponding $\Gamma$=2.01$\pm$0.08 is found to phenomenologically describe the high energy emission of HE 0436-4717.
	Among the different observations the photon index is consistent with being constant, and a lower limit E$_{\rm{cut-off}}>$280 keV is obtained. We tested a few
	Comptonization models, obtaining a lower limit of kT$_{\rm{e}}>$65 keV for the hot corona temperature.
	This temperature allowed us to estimate the optical depth for the HE 0436-4717 hot corona $\tau_{\rm{e}}<$1.3~.
	\item A narrow and constant \ion {Fe} K$\alpha$ emission line is observed, while a broader component is not required by these data. 
	Both the line and the associated Compton reflection component are in agreement with a scenario in which they arise from Compton-thick matter located far away from the central BH.

\end{itemize}

	
	
	\begin{acknowledgements} We are grateful to the referee whose comments improved the quality of this work.
		This  work  is  based  on  observations  obtained  with:  the NuSTAR mission,  a  project  led  by  the  California  Institute  of  Technology,  managed  by  the  Jet  Propulsion  Laboratory  and  funded  by  NASA; XMM-Newton,  an  ESA  science  mission  with  instruments  and  contributions  directly funded  by  ESA  Member  States  and  the  USA  (NASA). This  research  has  made  use  of  data,  software  and/or  web tools obtained from NASA’s High Energy Astrophysics Science  Archive  Research  Center  (HEASARC),  a  service  of Goddard  Space  Flight  Center  and  the  Smithsonian  Astrophysical  Observatory. Part of this work is based on archival data, software or online services provided by the Space Science Data Center - ASI.
		SB, RM and FV acknowledge financial support from ASI under grants ASI-INAF
		I/037/12/0 and n. 2017-14-H.O.
		F.T. acknowledges support by the Programma per Giovani Ricercatori - anno 2014 ``Rita Levi Montalcini''.
		AM acknowledges financial support from the Italian Space Agency under grant ASI/INAF I/037/12/0-011/13.
	\end{acknowledgements}
	
	\thispagestyle{empty}
	\bibliographystyle{aa}
	\bibliography{middeiHE04.bib}

	
\end{document}